\documentstyle[aps,epsfig,twocolumn,psfig19,floats]{revtex}
\renewcommand {\baselinestretch}{1.}

\newcommand{\aleq}{\mbox{\ 
\raisebox{-.9ex}{$\stackrel{\textstyle<}{\sim}$}\ }}
\newcommand{\ageq}{\mbox{\
\raisebox{-.9ex}{$\stackrel{\textstyle >}{\sim}$}\ }}

\def\u{{\mbox{\boldmath$u$}}}

\def\k{{\mbox{\boldmath$k$}}}

\def\la{{\langle}}
\def\ra{{\rangle}}

\def\eps{{\epsilon}}

\def\rel{{Re_\lambda}}

\def\begineq{\begin{equation}}
\def\endeq{\end{equation}}

\begin{document}
\bibliographystyle{prsty}

\title{Developed turbulence: \\ From full simulations to full mode
reductions}
\author{Siegfried Grossmann,
Detlef Lohse, and
Achim Reeh}
\address{
Fachbereich Physik der Universit\"at Marburg,
Renthof 6, D-35032 Marburg, Germany}

\date{\today}

\maketitle
\begin{abstract}
Developed Navier-Stokes turbulence is simulated 
with varying wavevector mode reductions.
The flatness and  the skewness
of the velocity derivative
depend on the degree of mode
reduction.
They show a crossover towards the value
of the full numerical simulation when the viscous subrange starts to be
resolved.
The
intermittency corrections of the scaling exponents $\zeta_p$ of the
p$^{th}$ order
velocity structure functions 
seem to depend mainly on the proper resolution of the inertial subrange.
{\it Universal}
scaling properties
(i.e., independent of the degree of mode reduction)
are found for the relative scaling
exponents $\rho_{p,q} =
(\zeta_p - \zeta_3 p/3 )/
(\zeta_q - \zeta_3 q/3 )$.
\end{abstract}



Even today fully developed turbulence is hard to access through full
numerical simulations of the Navier-Stokes equations because the number of
degrees of freedom increases with the
Taylor-Reynolds number roughly as $Re_\lambda^{9/2}$ \cite{my75,fri95}.
Consequently, models and approximations of the Navier-Stokes
dynamics with a reduced number of degrees of freedom are considered.
Models which embody the cascade type structure of turbulence
enjoyed increasing popularity in recent years, e.g., the so
called GOY model \cite{gle73,kad95}. Closer to the
Navier-Stokes dynamics is its reduced wavevector set approximation (REWA) 
\cite{egg91a,gnlo92b,gnlo94a,gro95b,uhl96}. REWA uses 
a reduced, geometrically scaling
subset of wavevectors on which the Navier-Stokes equation is solved. Very high
Taylor-Reynolds numbers up to $Re_\lambda = 7\cdot 10^4$
\cite{gnlo94a,gro95b} can be achieved. 

However, a
priori it is not clear whether these models and approximations are in the same
universality class as the Navier-Stokes dynamics itself,
as small scale structures corresponding to the high $k$ modes are not be fully
resolved. If inertial subrange (ISR)
scaling properties depend on details of the
viscous subrange (VSR) as speculated 
 for the GOY model \cite{sch95,lev95}, a cascade type
approach towards fully developed turbulence may not give the correct inertial
range scaling properties.
Moreover -- and as we will see more importantly --
 in these models the phase space has a different representation
than in 3D Navier-Stokes turbulence. 
Indeed, detailed REWA calculations
\cite{gnlo94a,uhl96}
for the
scaling exponents $\zeta_p$ of the
$p^{th}$ order longitudinal velocity structure functions
\begin{equation}
  D_i^{(p)} (r) =
  \langle ({\mbox{$u_i$}} ({\mbox{\boldmath$x$}}
  + r{\mbox{$\boldmath e_i $}} ) - {\mbox{$u_i$}}
  ({\mbox{\boldmath$x$}} ))^p \rangle \propto r^{\zeta_p},
\label{eq1}
\end{equation}
$i=1,2,3$,
show much smaller (but non vanishing \cite{gnlo94a,uhl96})
deviations $\delta\zeta_p = \zeta_p - p/3$ from
their classical 
values $\zeta_p = p/3$ (``K41'')
 than those from experimental measurements
\cite{ans84,ben93b} or full numerical simulations
(for  Reynolds numbers up to $\rel \approx 210$)
\cite{vin91,cao96}. Also, the flatness
$F_i=\la (\partial_i u_i )^4\ra/
\la (\partial_i u_i )^2\ra^2$ is $\approx 3.15$ \cite{gnlo92b}
for all $\rel$ in contrast to
experiments and full simulations where it seems to increase \cite{sre96b}
with $\rel$.
Analogous results hold for the skewness 
$S_i=\la (\partial_i u_i )^3\ra/
\la (\partial_i u_i )^2\ra^{3/2}$.
On the other hand, REWA may well represent the ``correct'' large
$\rel \ageq 10^3 $ limit
where $F$ and $S$
are speculated to  become independent of $\rel$ \cite{tab96}.

In this letter we  systematically analyse how the
scaling properties change with an 
increasing degree of  wavevector mode reduction, i.e., we examine the
transition from full numerical simulations to reduced wavevector set
approximations.
Since full simulations are possible only for low $Re_\lambda$ values,  
the present calculations are restricted accordingly, even though  
REWA was constructed for the large $Re_\lambda$ limit.
There is at most a short ISR.
However, the  
extended self-similarity method (ESS) \cite{ben93b} allows us to extract  
scaling exponents.

The aim of the work is to better understand the origin of intermittency
scaling corrections. Two views are discussed: The meanwhile classical
multifractal picture (see e.g.\ ref.\ \cite{fri95} for a review)
in which intermittent fluctuations build
up in the ISR and the Leveque-She reflection picture \cite{lev95} in which ISR
quantities depend on VSR properties. The importance of the latter mechanism
could be shown for the GOY model \cite{lev95,sch95}. However, 
for 3D Navier-Stokes
turbulence with different kind of hyperviscosity \cite{cao96},
no dependence of the ISR 
scaling exponents $\zeta_p$ on the kind of
hyperviscosity
could be detected.
Our analysis seems to support this result.
VSR effects on
$\zeta_p$ could not be identified.
Our interpretation is 
that the proper {\it local}
phase space resolution is of prime importance
for the correct representation of
the scaling corrections $\delta\zeta_p$.

We now describe our  analysis in detail.
The 3D incompressible
Navier-Stokes equations are numerically solved on a $N^3$
grid with periodic boundary conditions.
Spherical truncation is used to reduce aliasing. 
We force the system on the largest scale
(wavevectors $\k = (0,0,1)/L$ and permutations thereof)
as e.g.\ described in ref.\
\cite{gnlo94a}. Units are fixed by picking the length scale $L=1$ and the
average energy input rate (= the energy dissipation rate) $\eps=1$.
The Taylor-Reynolds number is defined as
 $\rel= u_{1,rms}\lambda /\nu$,
 where $\lambda = u_{1,rms}/ (\partial_1 u_1)_{rms}$ is the Taylor
 length and 
 $\nu$  the viscosity.
Our results refer to $N=60$ and $\nu =0.009$, corresponding to 
a resolution of scales  $r \ge 2\pi L/N \approx 3.6\eta$
and $\rel \sim 100$.
Time integrations
of about 
60 large eddy turnover times are performed.
Averages are taken over space and
time.
We also did shorter runs for $N=80$ and longer runs for $N=48$ which gave the
same results.

As our key parameter we
now introduce the wavenumber $k_B$ with $1 < k_B \le k_{max}=N/2$,
characterizing the degree of mode reduction: For a simulation with given $k_B$
all wavevectors with $|\k | \le k_B$ and scaled replica $2^l \k$,
$l=1,2,3,\dots$, thereof are considered; the mode amplitudes of the
remaining wavevectors are put to
zero. The choice $k_B = k_{max} = N/2$ corresponds to
a full simulation,  $k_B\sim 2$ is 
our former REWA calculation
\cite{egg91a,gnlo92b,gnlo94a,gro95b}. For those calculations a pure
spectral code could be used; here, because of the huge increase of couplings, a
pseudospectral code as described in \cite{can88,vin91} was employed.

\begin{figure}[htb]
\setlength{\unitlength}{1.0cm}
\begin{picture}(6,6)
\put(0.5,0.5)
{\psfig{figure=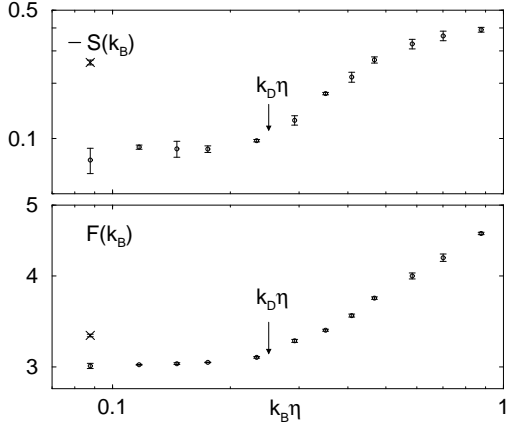,width=7cm,angle=-90}}
\end{picture}
\caption[]{
Skewness $S$ and flatness $F$ as  functions of $k_B$.
The error bars express the statistical differences of the values $S_i$
and $F_i$ for the three space directions $i=1,2,3$.
With the chosen lattice resolution $S$ and $F$ roughly reach their
saturation value.
The 
crosses on the very left refer to the REWA calculation with $k_B=3$,
but with a full VSR resolution for $k>9$.
}
\label{fig_fs}
\end{figure}

Figure \ref{fig_fs} shows the skewness $S$ and the flatness
$F$ as a function of
$k_B$. 
A crossover at
$k_D \approx
 1/(4\eta)\approx 9 $ can be identified. 
Here, $k_D$ denotes the wave number with maximal dissipation rate,
where massive viscous damping starts in the spectrum.
For $k_B < k_D$ the flatness and  the skewness
essentially remain on their REWA values. But at $k_B>k_D$
they start to drastically increase towards their saturated values corresponding
to the full simulation.

\begin{figure}[htb]
\setlength{\unitlength}{1.0cm}
\begin{picture}(6,11.5)
\put(0.5,6.5)
{\psfig{figure=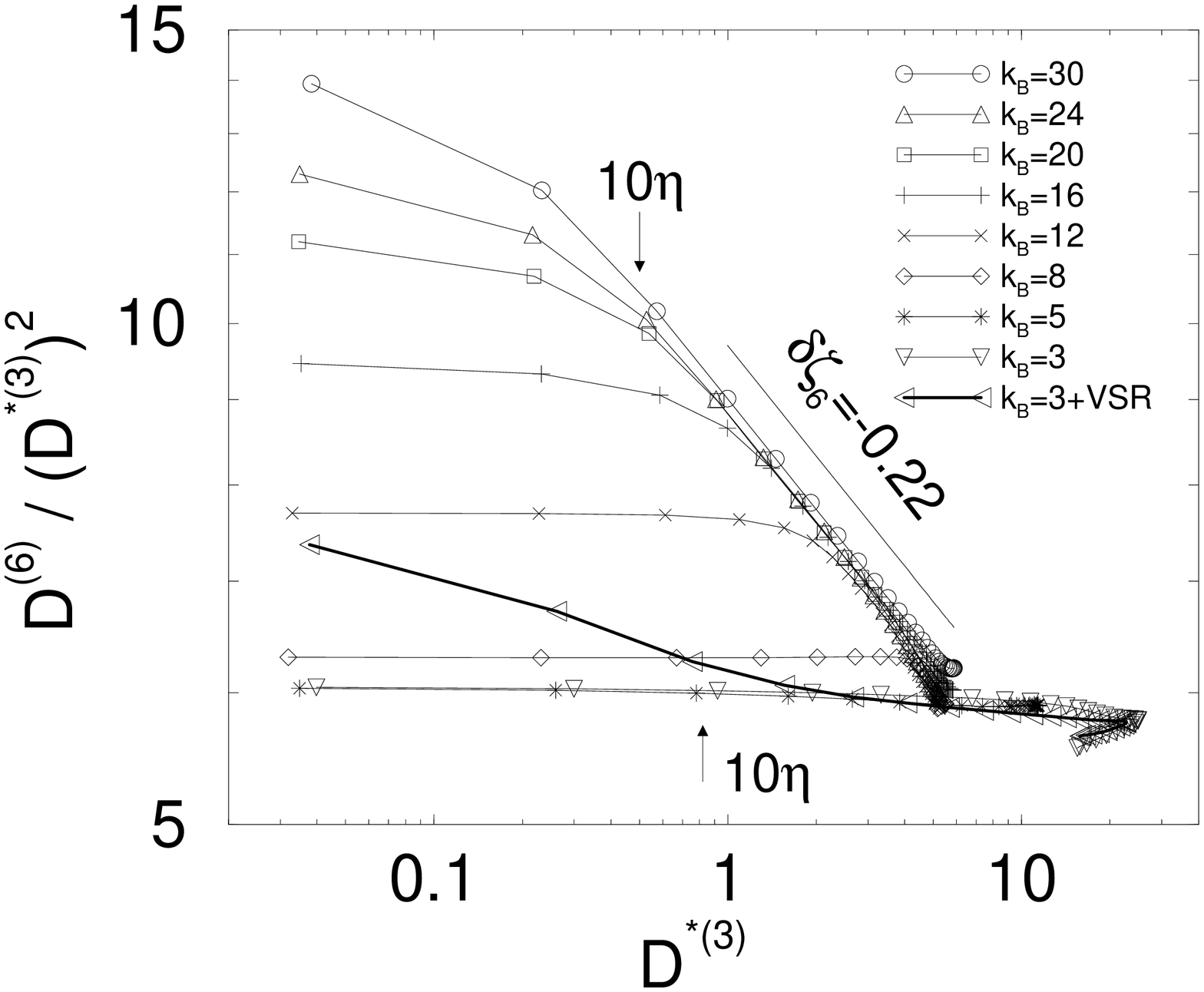,width=7cm,angle=-90}}
\put(0.5,0.5)
{\psfig{figure=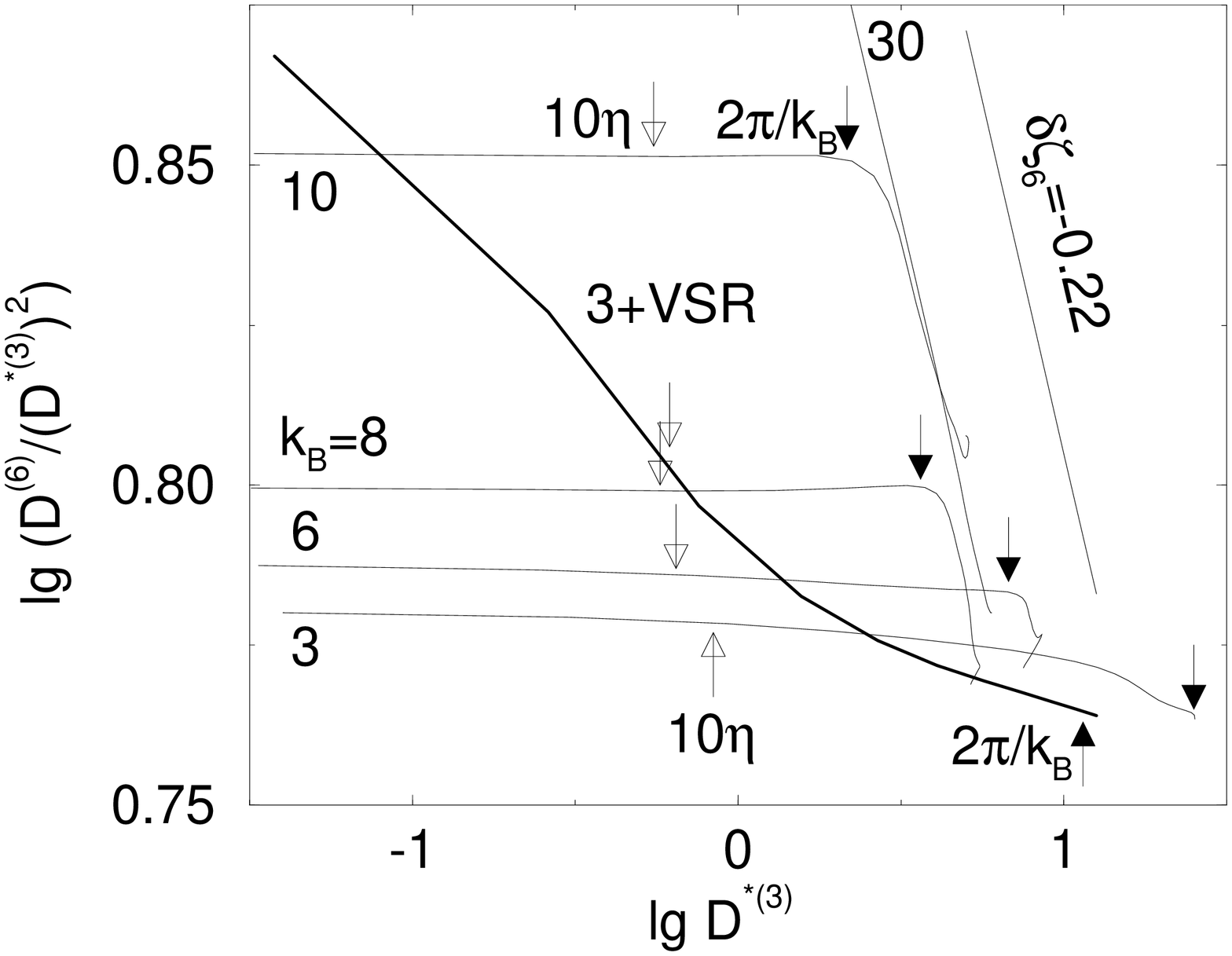,width=7cm,angle=-90}}
\end{picture}
\caption[]
{
(a) Compensated  ESS type plot
$D^{(6)}(r)/(D^{*(3)}(r))^2$ vs $D^{*(3)}(r)$
for various $k_B$.
The slopes in this type of figure are the $\delta\zeta_p$'s;
a horizontal line thus means K41. 
The open arrows point towards
$r=10\eta$ and the filled ones to $r=2\pi/k_B$. The outer length scale
$L$ is the very last point of each curve. 
Figure (b) shows an enlargement of (a) which illustratively
shows the scaling with $\zeta_6 = -0.22$ in the $[2\pi/k_b, L] $ regime.
}
\label{fig_ess}
\end{figure}

Figure \ref{fig_ess}
shows the
{\it compensated}
structure function $D^{(6)}(r)/[D^{*(3)}(r)]^2$
vs $D^{*(3)}(r)$. This kind of plot
allows for a better detection of local {\it deviations}
from scaling than the standard ESS \cite{ben93b} plot
$D^{(6)}$ vs $D^{*(3)}$. 
We find that for $k_B \ge 5$ the value $\delta\zeta_6 \approx
- 0.22$ is always
a good fit in the large $r$ regime between $2\pi /k_B$ and $L$.
This scaling regime shrinks for decreasing $k_B$ and
vanishes below $k_B\approx 5$ as then $2\pi /k_B$
 essentially collapses with the
external length scale $L$.

Figure \ref{fig_ess} 
suggests that at least for
small $\rel$ for $10\eta \ll 2\pi/k_B \ll L$
(a condition which never is really reached in our small $\rel$ simulations;
the simulation for $k_B=6$ is closest to it, see in particular
figure \ref{fig_ess}b)
there are three ranges: The (underresolved)
VSR $r\ll 10\eta$ where of
course $D^{(6)} \propto (D^{(3)})^2$,
a REWA ISR in the underresolved regime $[10\eta, 2\pi/k_B]$
with very small but nonvanishing (note the nonzero slope in figure
\ref{fig_ess}b in that regime)
intermittency corrections \cite{gnlo94a,uhl96}, and the
fully mode resolved Navier-Stokes ISR $[2\pi/k_B , L]$ with 
the intermittency corrections $\delta\zeta_6 = - 0.22$ as in full numerical
simulations.
This prompts the conclusion that
it is the {\it local} phase space resolution and not a proper
VSR resolution which is essential for the correct representation of scaling
corrections.

To further support this statement,
we performed a simulation with $k_B =3$,
but in addition a full resolution of all
modes $k > 9$, i.e., of the VSR. This curve is also shown in figure
\ref{fig_ess}. Indeed, there are hardly any scaling corrections in the ISR,
$\delta\zeta_6 \approx 0$. On the other hand,
as expected both the flatness and the
skewness are much bigger than in the REWA calculations \cite{gnlo92b}
as now the VSR is better resolved. We added the corresponding data points
in figure \ref{fig_fs}.

As is well known, the Navier-Stokes intermittency corrections 
are well fitted by the She-Leveque (SL)
model \cite{she94,fri95,che95,she95}
\begineq
\zeta_p = {p\over 3} - C_0 \left( {p\over 3} (1-\beta^3 ) - (1-\beta^p )
\right). 
\label{eq6}
\endeq
with the parameters $\beta$ and $C_0$, which in ref.\ \cite{she94} where
suggested to be $\beta = (2/3)^{1/3}$ and $C_0 = 2$.  In eq.\ (\ref{eq6}) we
already used the restriction $\zeta_3 = 1$ to eliminate a third parameter which
was introduced
in the original work \cite{she94,she95}. The parameter $C_0$
was related to 
the rate at which the probability to find the most intermittent events
decays in the large $k$
limit and also interpreted as the 
the codimension of the dissipative structures
\cite{she94,she95,lev95}. If in 
3D Navier-Stokes turbulence these are 1D filaments, we have
$C_0=2$.
This interpretation also works for REWA turbulence ($k_B=3$)
\cite{egg91a,gnlo94a}: The dissipative structures are nearly
3D because of the
lack of large wavevector resolution. Therefore, $C_0\approx 0$ and
according to (\ref{eq6}) $\zeta_p \approx p/3$, in agreement with the
numerical results \cite{gnlo94a,uhl96}. However, the interpretation
of $C_0$ as codimension of the dissipative structures seems to be at variance
with the simulation with $10\eta \ll 2\pi /k_B \ll L$, where we find the 3D
Navier Stokes values for $\delta\zeta_p$ in the $[2\pi/k_B ,
L]$ regime in spite
of the poor VSR resolution which determines the dimension of the dissipative
structures.

\begin{figure}[htb]
\setlength{\unitlength}{1.0cm}
\begin{picture}(6,6)
\put(0.5,0.5)
{\psfig{figure=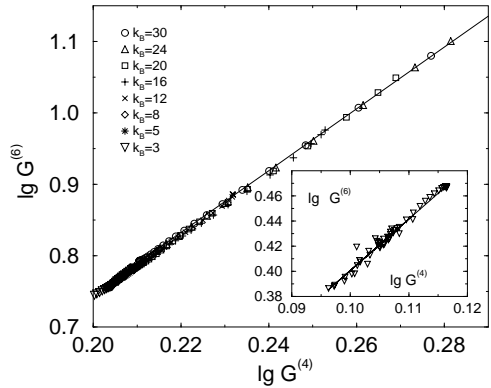,width=7cm,angle=-90}}
\end{picture}
\caption[]{
GESS type plot 
$G^{(6)}(r)$ vs $G^{(4)}(r)$ for various $k_B$.
The inset focuses on the REWA scaling range for much larger $\rel
\approx 7\cdot 10^4$  \cite{gro95b}.
The $G$-functions in the inset are calculated from the {\it total}
structure functions, not from the longitudinal ones as in the main figure.
Even for this huge Reynolds number the scaling range 
is very short because of the small intermittency corrections
$\delta\zeta_p$. Note that in this type of plot the whole VSR collapses
into the upper right point of the curves.
}
\label{fig_gess}
\end{figure}

Whereas the scaling corrections $\delta\zeta_p$ do depend on the local phase
space resolution, their ratios
\begineq
\rho_{p,q} = {\zeta_p - \zeta_3 p/3 \over \zeta_q -\zeta_3 q /3}
\label{eq2}
\endeq
do not. These exponents are 
the relative scaling exponent of two compensated structure functions
\cite{ben95}
\begineq
G^{(p)} (r) = D^{(p)} (r) / (D^{*(3)}(r))^{p/3}.
\label{eq3}
\endeq
Indeed, 
the scaling of $G^{(p)}$ vs $G^{(q)}$ is perfect
in the whole range we resolve, as seen from
the so called generalized ESS (=GESS, \cite{ben95}) figure \ref{fig_gess}.
The reason is that in the GESS plot all data points of the VSR
{\it collapse} since $G^{(p)}(r)$ is constant for $r \aleq 10 \eta$. 
The $\rho_{p,q}$ exponents could be obtained from plots like that
in figure \ref{fig_gess}. Alternatively, they can be obtained
via the $\delta\zeta_p$ from
straight line fits to compensated
ESS type plots as in figure \ref{fig_ess}
which turns out to result in a smaller statistical error. 
Though obviously the $\delta\zeta_p$ in figure \ref{fig_ess} depend on the
chosen $D^{*(3)}$ range, their ratios $\rho_{p,q}$ do {\it not} depend on this
range.

\begin{figure}[htb]
\setlength{\unitlength}{1.0cm}
\begin{picture}(5,4.0)
\put(0.1,1.0)
{\psfig{figure=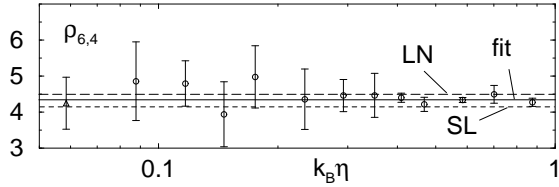,width=8cm,angle=-90}}
\end{picture}
\caption[]{
Relative scaling exponent $\rho_{6,4}$ as a function of
our mode reduction parameter $k_B$. Here, $\eta$ is the Kolmogorov inner
scale.
The error bars take care of both the quality of the linear
regression of the scaling laws and the deviations from isotropy.  
The triangle on the very left  refers to the REWA calculation
from ref.\ \cite{gro95b} 
for much larger $\rel$.
The dotted line shows
the SL prediction \cite{she94} from equation
(\ref{eq5}) with $\beta = (2/3)^{1/3} \approx 0.8736$, i.e.,
$\rho_{6,4}=4.14$, the dashed one the prediction
from the LN model $\rho_{6.4} = 4.5$, and the solid line is the best
fit 
$\rho_{6,4}= 4.33\pm 0.05$, implying 
$\beta= 0.94 \pm 0.02 $ in the SL mode, very close to the value
$\beta = (7/9)^{1/3} \approx 0.92$ suggested in ref.\ \cite{che95}.
For different pairs $(p,q)$ the universality and the agreement with
SL and LN is correspondingly good.
}
\label{fig_rho}
\end{figure}

{}From figure \ref{fig_rho} we see that the $\rho_{p,q}$ also
do not depend on
$k_B$. Similar universality has recently been found by Benzi et al.\
\cite{ben96c} for the GOY shell model where the $\zeta_p$
depend on the spacing between the shells, but the $\rho_{p,q}$ do not.

The relative scaling exponents $\rho_{p,q}$ are rather well described both by
the SL model, predicting
\begineq
\rho_{p,q} = {
(1-\beta^{p}) - (p/3)(1-\beta^3 ) \over
(1-\beta^{q}) - (q/3)(1-\beta^3 )},
\label{eq5}
\endeq
which only depends on $\beta$ and not on the codimension of
the dissipative structures $C_0$ any more,
and by the log normal model (LN, see ref.\ \cite{fri95} for
a detailed discription), predicting
\begineq
\rho_{p,q} = {p(p-3)\over q (q-3)}.
\label{ln}
\endeq
Our values are roughly $6\%$ below those from the LN model and
$6\%$ above the SL prediction with the originally suggested
$\beta = (2/3)^{1/3}$. By taking the suggestion of Chen and Cao \cite{che95},
$\beta = (7/9)^{1/3}$, the agreement can be improved.

Our results for both the skewness $S$, the flatness $F$,
and for the scaling corrections
$\delta\zeta_p$ are consistent with Cao et al.'s \cite{cao96} full numerical
Navier-Stokes simulations for normal and hyperviscous damping
$(-)^h \nu \nabla^{2h} \u$ ($h=1$
 means normal viscosity).
 Our large wavevector reduction is similar to a kind of hyperviscosity.
 Indeed, the
larger $h$
the smaller the flatness is. As in the calculations presented here,
the ISR scaling properties $\delta\zeta_p$
are hardly affected by the kind
of hypervisocity.

In summary, we repeat that
in 3D Navier-Stokes turbulence
the main origin of intermittency corrections
seems to be the proper resolution of the phase space at the scale
of interest. Reflections from the VSR seem to be of minor importance.
How come that the energy flux reflected from the VSR
\cite{lev95,sch95}
reaches so far in the ISR for the GOY model but apparently not
for 3D Navier-Stokes turbulence?
We speculate that in 3D the phases of the modes are
 subjected to far more
fluctuations than in the 1D GOY model. Therefore, coherences
get destroyed easier. Some coherence however must remain, otherwise
no energy could be transported downscale.

\noindent
{\bf Acknowledgements:}
Very helpful discussion with
L.\ Biferale, J.\ Eggers, L.\ Kadanoff, Z.S.\ She, and K.\ R.\
Sreenivasan are acknowledged.   
We in addition
take great pleasure in expressing our gratitude to Leo Kadanoff
for all he has taught us through his penetrating physical insight
and originality, for his enduring support, and for his warm friendship.
Our pleasure is doubled by having the opportunity to congratulate him
on the occasion of his sixtieth birthday.  Support for this work by
the Deutsche Forschungsgemeinschaft (DFG) under grant SBF185 and by
the German-Israel Foundation (GIF) is also gratefully acknowledged.

\vspace{0.5cm}

\noindent
 e-mail addresses:\\
grossmann$\_$s@physik.uni-marburg.de\\ 
lohse@physik.uni-marburg.de\\
reeh@mailer.uni-marburg.de


\begin{thebibliography}{10}

\bibitem{my75}
A.~S. Monin and A.~M. Yaglom, {\em Statistical Fluid Mechanics} (The MIT Press,
  Cambridge, Massachusetts, 1975),
M. Nelkin, Advances in Physics {\bf 43},  143  (1994).

\bibitem{fri95}
U. Frisch, {\em Turbulence} (Cambridge University Press, Cambridge, 1995).

\bibitem{gle73}
E.~B. Gledzer, Sov. Phys. Dokl. {\bf 18},  216  (1973);
M. Yamada and K. Ohkitani, J. Phys. Soc. Jpn. {\bf 56},  4210  (1987);
Prog. Theor. Phys. {\bf 79},  1265  (1988);
M.~H. Jensen, G. Paladin, and A. Vulpiani, Phys. Rev. A {\bf 43},  798
(1991);
D. Pisarenko {\it et~al.}, Phys. Fluids A {\bf 5},  2533  (1993).

\bibitem{kad95}
L. Kadanoff, D. Lohse, J. Wang, and R. Benzi, Phys. Fluids {\bf 7},  617
  (1995).

\bibitem{egg91a}
J. Eggers and S. Grossmann, Phys. Fluids A {\bf 3},  1958  (1991).

\bibitem{gnlo92b}
S. Grossmann and D. Lohse, Z. Phys. B {\bf 89},  11  (1992).

\bibitem{gnlo94a}
S. Grossmann and D. Lohse, Phys. Fluids {\bf 6},  611  (1994);
Phys. Rev. E {\bf 50},  2784  (1994).

\bibitem{gro95b}
S. Grossmann, D. Lohse, and A. Reeh,  in {\em Dynamical Systems and Chaos, Vol
  2: Physics}, edited by S.~Saito und K.~Shiraiwa Y.~Aizawa (World Scientific,
  Singapore, 1995), p.\ 209.

\bibitem{uhl96}
C. Uhlig and J. Eggers, Preprint  (1996).

\bibitem{sch95}
N. Sch\"orghofer, L. Kadanoff, and D. Lohse, Physica D {\bf 88},  40  (1995);
L.~P. Kadanoff, Physics Today {\bf 48},  11  (1995).

\bibitem{lev95}
E. Leveque and Z.~S. She, Phys. Rev. Lett. {\bf 75},  2690  (1995).


\bibitem{ans84}
F. Anselmet, Y. Gagne, E.~J. Hopfinger, and R.A. Antonia, J. Fluid Mech. {\bf
  140},  63  (1984);
Ch. Meneveau and K.~R. Sreenivasan, J. Fluid Mech. {\bf 224},  429  (1991);
A.~Arneodo et~al., Europhys. Lett. {\bf 34},  411  (1996);
F. Belin, P. Tabeling, and H. Willaime, Physica D {\bf 93},  52  (1996).

\bibitem{ben93b}
R. Benzi {\it et~al.}, Phys. Rev. E {\bf 48},  R29  (1993).

\bibitem{vin91}
A. Vincent and M. Meneguzzi, J. Fluid Mech. {\bf 225},  1  (1991);
Z.~S. She {\it et~al.}, Phys. Rev. Lett. {\bf 70},  3251  (1993).


\bibitem{cao96}
N. Cao, S. Chen, and Z.~S. She, Phys. Rev. Lett. {\bf 76},  3711  (1996).

\bibitem{sre96b}
K.~R. Sreenivasan and R.~A. Antonia, Annual Review of Fluid Mech. {\bf x},  y
  (1996).

\bibitem{tab96}
P. Tabeling {\it et~al.}, Phys. Rev. E {\bf 53},  1613  (1996);
V. Emsellem {\it et~al.}, Phys. Rev. E {\bf x},  y  (1996).

\bibitem{can88}
C. Canuto, M.~Y. Hussaini, A. Quarteroni, and T.~A. Zang, {\em Spectral Methods
  in Fluid Dynamics} (Springer Verlag, Heidelberg, 1988).

\bibitem{she94}
Z.~S. She and E. Leveque, Phys. Rev. Lett. {\bf 72},  336  (1994).

\bibitem{che95}
S. Chen and N. Cao, Phys. Rev. E {\bf 52},  R5757  (1995).

\bibitem{she95}
Z.~S. She and E.~S. Waymire, Phys. Rev. Lett. {\bf 74},  262  (1995);
B. Dubrulle, Phys. Rev. Lett. {\bf 73},  959  (1994).

\bibitem{ben95}
R. Benzi {\it et~al.}, Europhys. Lett. {\bf 32},  709  (1995);
Physica D {\bf 1317},  1  (1996).

\bibitem{ben96c}
R. Benzi, L. Biferale, and E. Travatore, Phys. Rev. Lett. {\bf 77},  3114
  (1996).


\end{thebibliography}

\end{document}